\documentstyle[11pt]{article}
\title{\begin{flushright}
{\normalsize TPI-MINN-93/22-T \\
NUC-MINN-93/10-T \\
BUTP-93/13 \\
May 1993 \\}
\end{flushright}
\bf SCREENING MASS FROM CHIRAL\\ PERTURBATION THEORY, VIRIAL\\
EXPANSION, AND THE LATTICE }
\author{{\bf V. L. Eletsky}$^{\dagger}$ \\
  {\small\it Institute for Theoretical Physics, Bern University,
  CH-3012 Bern, Switzerland}
 \and
  {\bf J. I. Kapusta} \\
  {\small\it School of Physics and Astronomy,
  University of Minnesota, Minneapolis, MN 55455}
 \and
{\bf R. Venugopalan} \\
  {\small\it Theoretical Physics Institute,
  University of Minnesota, Minneapolis, MN 55455}}

\date{}

\begin{document}

\maketitle

\begin{center}
Abstract\\
\end{center}

We calculate the electric screening mass in hot hadronic matter using
two different approaches, chiral perturbation theory and the relativistic
virial expansion with empirical phase shifts, and compare the results
to each other and to a gas of free pions and $\rho$ mesons.  We also
compute the electric screening mass for noninteracting,
charged bosons with mass $m$ on a lattice to study likely finite size
effects in lattice gauge theory simulations of continuum QCD.  For
a lattice of given size, the continuum can be properly represented
only for a window in the ratio $T/m$.

\vfill \eject

\section{Introduction}

The advent of a new generation of heavy ion accelerators is making it
possible to produce hadronic matter at high temperatures and densities.
Specifically, we refer to temperatures near the pion mass and densities
greater than the density at the center of atomic nuclei.
Little is known about the properties of
these hot and dense systems and of their expected transition to a phase
consisting of quarks and gluons. In principle, both of these phases are
described by QCD, but the structure of this theory is especially complicated
for
the above mentioned region of temperatures and densities. It is therefore
reasonable to use a wide array of techniques to investigate different
aspects of the behavior of hot and dense hadronic matter.

One way in which we can investigate the properties of this many-body system
is to study its response to a small perturbation using linear response
theory. This response can then be expressed in terms of correlation functions
unperturbed by the presence of the probe.  For example, we could ask
for the response of an electrically neutral system to an applied, static,
electric field. The system responds to a weak perturbation, such as
a heavy, charged lepton or hadron, by dynamically Debye screening the long
range Coulomb force.  The Debye screening length is independent of the
external perturbation.

In Ref.~\cite{irv} the inverse of the Debye screening length, the electric
screening mass, $m_{el}$, was studied in hot QCD. Interestingly, this can
be done in two different ways. One way is to compute the photon self-energy.
The electric screening mass squared is then the static,
infrared limit of the time-time component of the self--energy. Another way
is through an identity which relates the electric screening
mass squared to the pressure
\begin{equation}
m_{el}^2=e^2\,\left(\frac{\partial^2 P}{\partial\mu^2}\right)_{\mu =0}
\, ,
\label{mel}
\end{equation}
where $\mu$ is the electric charge chemical potential.
Both ways of computing the screening mass are exact.

In this work, we will use Eq.~(\ref{mel}) to calculate the electric screening
mass for a hot pion gas. This is done using three very different techniques,
namely, the relativistic virial expansion, chiral perturbation theory,
and lattice theory, which are discussed in successive sections.
In the virial approach, dynamical information obtained from empirical
two-body scattering phase shifts is used to compute the pressure for an
interacting pion gas with a non-zero chemical potential and hence, from
Eq.~(\ref{mel}), the electric screening mass. In the following section the
screening mass is computed using finite temperature chiral perturbation
theory, extended to the case of a non-zero chemical potential.
The results of chiral perturbation theory are found to agree
exactly with those of the virial expansion in the low temperature Boltzmann
limit, $T<<m_\pi$, to order $(T/m_\pi)^{3/2}$. Finally, we compute the
screening mass for free, massive, bosons on a lattice. The ratio of the
screening mass on the lattice to that in the continuum is studied
parametrically as a function of $T/m$, where $m$ is the mass of the boson.
This gives an indication of how large a lattice is needed in order that
lattice gauge theory properly approach the continuum QCD limit.
See Fig. 1, taken from Ref.~\cite{irv}.

Each of the above mentioned methods, as might be expected, has its advantages
and disadvantages. The relativistic virial expansion demonstrates how the
influence of both resonant and repulsive interactions may be included in
calculations of the electric screening mass at
temperatures close to the pion mass. Since the virial expansion can be
expressed as an expansion in powers of the density, the results of this
approach very likely contain the right physics for dilute systems. If, however,
the system is dense, three- and higher-body interactions are significant.
Extracting this information from the empirical phase shifts is difficult. The
chiral perturbation theory approach is of interest since it contains many of
the low energy properties of QCD. It also explains some of the low energy
hadron phenomenology successfully.  Furthermore, since the Lagrangian is
known, many quantities of physical interest can be studied.
A limitation of this approach, though, is that resonant interactions
are not fully accounted for.  These may be expected to
contribute significantly for temperatures close to the pion mass. Currently,
lattice gauge theory is a  popular technique to understand the structure
of strongly interacting matter. This includes studies of various correlation
functions of mesons on the lattice. While lattice gauge theory is in
principle very powerful, finite size effects are important.
Analytic calculations for free, massive, bosons on the lattice are therefore
very useful in quantifying the sizes of these effects.

Besides being of intrinsic physical interest, our calculation of the electric
screening mass is also illustrative because these techniques may be used
to compute dispersion relations and other response functions.
The importance of
alternative techniques to calculate various correlation functions
has been discussed by Shuryak~\cite{Shyk}, who has used both experimental
phase shifts as well as QCD sum rule techniques to compute
dispersion relations for hot hadronic matter~\cite{Shyk2}.
In this way we may get
a better physical understanding of the immense information that may
in principle be available from both heavy ion experiments and future
lattice gauge theory simulations, especially with the Teraflop project.

\section{Relativistic Virial Expansion}

The relativistic virial expansion has been used recently to compute the
thermodynamic properties of a dilute gas of interacting hadrons
\cite{wvp,rvmp}.  This approach is appealing because it allows one to
systematically include the effects of both resonant and repulsive interactions
in finite temperature hadronic matter by relating the state variables
to the known, empirical, phase shifts.
In this section we will use the relativistic virial expansion to
compute the electric screening mass for a gas of interacting pions. It will be
shown in the following section that the virial expansion results also
provide an excellent check of chiral perturbation theory calculations
with nonzero chemical potentials.

The relativistic virial expansion, introduced by Dashen and
co--workers \cite{Dash}, relates thermodynamic state variables of a system of
interacting particles to the ${\cal S}$--matrix. The partition function is
separable into a product of the non--interacting partition function $Z_0$ and
an interacting term which is proportional to bi--linear products of the
$S$--matrix and its inverse. The logarithm of the partition function can be
written as
\begin{eqnarray}
\ln Z = \ln Z_0 + \sum_{i_{1},i_{2}} z_1^{i_1} z_2^{i_2} b(i_{1},i_{2}) \, ,
\label{prvr}
\end{eqnarray}
where $z_j = \exp (\beta \mu_j)$ for $j=1,2$ are the fugacities.
The virial coefficients $b(i_{1},i_{2})$ are written as
\begin{eqnarray}
 b(i_{1},i_{2})&=&\frac{V}{4\pi{i}}\int\frac{d^{3}P}{(2\pi)^{3}}
\int dE\exp\left(-\beta(P^{2}+E^{2})^{1/2}\right)\,\nonumber \\
&\times& {\rm
Tr}_{i_1,i_2} \big[A{\cal
S}^{-1}(E)\frac{\stackrel{\leftrightarrow}{\partial}}{\partial E}
{\cal S}(E)\big]_{c} \, .\label{form}
\end{eqnarray}
In the above, the inverse temperature is denoted by $\beta$ while $V$, $P$ and
$E$ stand for the volume, the total center of mass momentum and energy,
respectively. The labels  $i_1$ and $i_2$ refer to a channel of the ${\cal
S}$--matrix which has an initial state containing $i_1+i_2$
particles--the trace is therefore over all combinations of particle number. The
symbol $A$ denotes the symmetrization (anti--symmetrization) operator for a
system of bosons (fermions) while the expression with the double--sided arrow
is defined as
\begin{eqnarray}
{\cal S}^{-1} \left(
\stackrel{\leftrightarrow}{\partial}/\partial E \right){\cal S} \equiv
{\cal S}^{-1}(\partial{\cal S}/{\partial E}) -  (\partial{{\cal
S}^{-1}}/{\partial E}) {\cal S}\, .\label{coeff}
\end{eqnarray}
The subscript $c$ refers to the trace over all connected diagrams.
The lowest virial coefficient $B_2\equiv b(i_1,i_2)/V$ as
$V \rightarrow \infty$ corresponds to the case where $i_1=i_2=1$.

At temperatures close to the pion mass, the system is assumed to be
sufficiently dilute for the hadrons to interact chiefly via elastic collisions.
This assumption greatly simplifies Eq.~(\ref{coeff}) since the ${\cal
S}$--matrix can be expressed in terms of the phase shifts $\delta_{l}^{I}$ as
\begin{eqnarray}
{\cal S}(E)=\sum_{l,I} (2l+1)(2I+1)\exp(2\delta_{l}^{I}) \, ,
\end{eqnarray}
where $l$ and $I$ denote the angular momentum and isospin,
respectively. The above assumption also implies that two--body interactions are
dominant relative to the three--body and higher terms. The two-body
interactions
are expressed via the second virial coefficient
\begin{eqnarray}
B_2 = \frac{1}{2\pi^{3}\beta}\int_{M}^{\infty}dE \, E^{2}
K_{2}(\beta E) \,{\sum_{l,I}}^{\prime} g_{l,I}\frac{\partial\delta_l^I(E)}{
\partial E} \, .
\end{eqnarray}
The factor $g_{l,I}\equiv (2l+1)(2I+1)$ is the degeneracy of the $(l,I)$
channel and $M$ is the invariant mass of the interacting pair at threshold. The
prime over the summation sign denotes that for given $l$ the sum over $I$ is
restricted to values consistent with statistics. If the phase shifts
$\delta_l^I \rightarrow 0$ as $E \rightarrow M$, an integration by parts
yields
\begin{eqnarray}
B_2 = \frac{1}{2\pi^3}\int_{M}^{\infty}dE \, E^2 K_{1}(\beta E)
{\sum_{l,I}}^{\prime} g_{l,I}\delta_{l}^{I} \,.
\end{eqnarray}
For further details on the behaviour of the virial coefficients and expressions
for the thermodynamic state variables, we refer the reader to Ref.~\cite{rvmp}.

In the pion gas, the $\pi^+$, $\pi^-$ and $\pi^0$ have chemical potentials
$+\mu$, $-\mu$ and zero, respectively. The chemical potential in
Eq.~(\ref{prvr}),
$\mu_1 + \mu_2 \equiv \mu_Q$, corresponds to the net conserved charge $Q$
in each scattering channel contributing to Eq.~(\ref{prvr}). The $\pi\pi$
pressure due to interactions can then be expressed as
\begin{eqnarray}
P_{\pi\pi}^{int}=\sum_Q P_Q^{int}\, ;\qquad P_Q^{int}=Te^{\beta\mu_Q}
 B_{2,Q} \,\,,
\end{eqnarray}
where for $-2\leq Q\leq 2$, $-2\mu\leq\mu_Q\leq 2\mu$. The different $\pi\pi$
channels contributing to the second virial coefficient $B_{2,Q}$ for each $Q$
can then be decomposed, with the appropriate Clebsch factors, into the
corresponding spin--isospin channels. After a little algebra, the interacting
$\pi\pi$ pressure in the above equation is finally written in terms of the
spin--isospin phase shifts as
\begin{eqnarray}
P_{\pi\pi}^{int}&=&\frac{T}{2\pi^3}\int_{2m_\pi}^{\infty} dE\,
E^2 K_1(\beta
E)\bigg[ 2\cosh(2\mu\beta)\delta_0^2 + 2\cosh(\mu\beta) (\delta_0^2+3
\delta_1^1)\nonumber \\
&+&\delta_0^2+3\delta_1^1+\delta_0^0\bigg] \, .
\label{pint}
\end{eqnarray}
The total pion pressure is given by the sum of the above interacting pressure
and the ideal gas pressure
\begin{equation}
P_{\pi}^{ideal}
=\frac{1}{6\pi^2}\int_0^\infty {dp\, p^4 \over \omega} \bigg[{1\over {
e^{(\omega-\mu)/T}-1}}+
{1\over {e^{(\omega+\mu)/T}-1}}+{1\over {e^{\omega/T}-1}}\bigg] \, .
\label{pideal}
\end{equation}

The electric screening mass for the interacting pion gas can then be obtained
from Eq.~(1).  We get
\begin{eqnarray}
m_{el}^2=\frac{e^2 m_\pi^2}{\pi^2}\left[ \sum_{n=1}^\infty K_2({\beta n m_\pi})
+{1\over {m_\pi^2 \pi T}}\int_{2m_\pi}^{\infty} dE E^2 K_1
(\beta E)
\left(5\delta_0^2+3\delta_1^1\right)\right] \, .\label{intmel}
\end{eqnarray}
The second term in the above equation is the contribution to the electric
screening mass from the interactions. In Ref.~\cite{wvp} it was shown that
the pressure of an interacting pion gas was nearly identically the sum of the
pressures of an ideal gas of pions and (Maxwellian) $\rho$ mesons up to
rather large values of the temperature. This was due to a near exact
cancellation of the isospin weighted sum of the spin--zero phase shifts
leaving only a contribution from the $\delta_1^1$--resonant phase shift.
Unlike the pressure, however, the screening mass for an interacting
$\pi\pi$ gas is {\it not} the
sum of the screening masses of an ideal gas of $\pi$ and $\rho$ mesons. In
Fig.~2 we plot the electric screening mass squared (in units of $e^2 T^2$)
for an ideal gas of $\pi$'s and $\rho$'s as well as for an interacting pion
gas using the expression in Eq.~(\ref{intmel}). They agree fairly well at lower
temperatures but disagree by $10$\% or more at temperatures above the pion
mass.
The screening masses in the two cases differ because of the presence of a
repulsive $\delta_0^2$ piece in the interacting $\pi\pi$ screening mass in
Eq.~(\ref{intmel}). In its absence, they would agree almost exactly for the
temperatures considered.

In the above formulae, the virial expansion has been truncated at the level of
the second virial coefficient; only two-body collisions were considered.
At temperatures above the pion mass, three- and higher-body interactions will
begin to contribute significantly.  Since it is virtually hopeless to
expect to obtain complete experimental data on the ${\cal S}$-matrix for
$m$--particles in and $n$--particles out,
the extension of this approach to high densities is limited.
One would have to rely on models of the many-body
interactions, as obtained from effective Lagrangians, for example.

\section{Chiral Perturbation Theory}

In this section, we will use the method of effective chiral Lagrangians
\cite{w,gal,gel} to
calculate the electrostatic screening mass in a pion gas at finite
temperature. We will generalize the calculation of the pressure of
a pion gas performed in Ref.~\cite{gel} to the case of a finite chemical
potential associated with electric charge and then obtain $m_{el}^2$ using
Eq.~(\ref{mel}).

The partition function is given by a Euclidean functional
integral
\begin{equation}
{\rm Tr}\, \exp (-H/T)=\int [dU] \exp\left (-\int_T d^4 x L_{eff} \right)
\label{z}
\end{equation}
where $U(x)=\exp (i\phi^{a}(x)\tau^{a}/F)$ is an $SU(2)$
matrix comprising the pion field
$\phi (x)$. The integration should be performed over all configurations which
are periodic in Euclidean time, $U({\bf x},x_4 +\beta)=U({\bf x}, x_4)$.
The effective Lagrangian $L_{eff}$ is expressed as an infinite set of terms
with increasing number of derivatives or quark masses,
\begin{equation}
L_{eff}=L^{(2)}+L^{(4)}+L^{(6)}+ \cdots \, .
\label{leff}
\end{equation}
The leading term corresponds to the nonlinear $\sigma$-model,
\begin{equation}
L^{(2)}=\frac{1}{4} F^2 {\rm Tr}\,\left(\partial_{\mu} U^{\dagger}
\partial_{\mu}U-
m_0^2(U+U^{\dagger})\right) \, .
\label{L2}
\end{equation}
The coupling constant $F$ is the pion decay constant in the chiral limit
and $m_0$ is the pion mass in the lowest order in quark masses,
\begin {equation}
m_0^2=-\frac{1}{2F^2} (m_u +m_d)<\bar{u}u+\bar{d}d> \, .
\label{m2}
\end{equation}
Neglecting isospin breaking in the quark masses, the next order
Lagrangian may be written in the following form \cite{gel},
\begin{eqnarray}
L^{(4)}&=&-\frac{1}{4}l_{1}\left[
{\rm Tr}\,(\partial_{\mu} U^{\dagger}\partial_{\mu}U) \right]^2-
\frac{1}{4}l_{2}{\rm Tr}\left(\partial_{\mu} U^{\dagger}\partial_{\nu}U
\right) {\rm Tr}\left(\partial_{\mu} U^{\dagger}\partial_{\nu}U\right)
\nonumber \\
& &\mbox{}+\frac{1}{8}l_{4}m_0^{2}\,
{\rm Tr}\left(\partial_{\mu} U^{\dagger}\partial_{\mu}U\right)
{\rm Tr}\left(U+U^{\dagger}\right) \nonumber \\ & &
-\frac{1}{16}(l_{3}+l_{4})m_0^{4}\left[ {\rm Tr}(U+U^{\dagger})\right]^2
 \, , \label{L4}
\end{eqnarray}
which involves four coupling constants $l_1,...,l_4$.
The quadratic $\phi$ terms in $L^{(2)}$ describe free mesons of mass
$m_0$. The terms of higher order in $\phi$ in $L^{(2)}$ and higher order
Lagrangians are considered as perturbations.

Inclusion of a finite chemical potential $\mu$ related to the electric
charge is equivalent to coupling the pions to an external constant
electromagnetic vector potential with only temporal component $A_4=-i\mu$
being non-zero and imaginary. This changes time derivatives in
Eqs. (\ref{L2}) and (\ref{L4}) to covariant ones:
\begin{equation}
\partial_{\alpha}\phi^{a}\rightarrow
D_{\alpha}^{ab}\phi^b =
(\partial_{\alpha}\delta^{ab}-\mu\delta_{\alpha 4}
\epsilon^{0ab})\phi^{b} \, .
\label{covd}
\end{equation}
This results in the corresponding frequency shifts in the imaginary time
propagators of charged pions \cite{Joe},
\begin{equation}
G_{\pm}=\frac{1}{(\omega_{n}\pm i\mu)^2+{\bf p}^2 +m_0^2} \, ,
\label{prop}
\end{equation}
where $\omega_{n}=2\pi Tn$. It is important that the same shift also occurs
in the vertices containing derivatives.

The pressure corresponds to the nonvacuum part of the
thermodynamic potential $\Omega$,
\begin{equation}
P(\mu)=\varepsilon_{0}-\Omega \, ,
\label{pmu1}
\end{equation}
\begin{equation}
\Omega=-\lim_{V\rightarrow\infty}\frac{T}{V}\ln[{\rm Tr}\,e^{-(H-\mu Q)/T}]\, ,
\end{equation}
where $\varepsilon_0$ is the zero temperature and chemical potential limit
of $\Omega$.
Thus, to calculate $P(\mu)$ one should consider all closed--loop diagrams
involving all possible couplings from $L_{eff}$. We will confine ourselves
here to the second order in the density of the pion gas and thus take
into account only the diagrams with one and two thermal loops .
First we consider the noninteracting pion gas thermodynamic
potential $\Omega_{\pm}^{(0)}$, and
pion propagators and their derivatives at the origin,
\begin{eqnarray}
\Omega_{\pm}^{(0)}&=&-\int\frac{d^3 p}{(2\pi)^3}T\sum_{n}
\ln (\omega^2+(\omega_{n}\pm i\mu)^2) \, , \\
G_{\pm}(0)&=&\int\frac{d^3 p}{(2\pi)^3}T\sum_{n}
\frac{1}{\omega^2+(\omega_{n}\pm i\mu)^2}  \, , \\
D_{0}G_{\pm}(0)&=&\int\frac{d^3 p}{(2\pi)^3}T\sum_{n}
\frac{i(\omega_{n}\pm i\mu)}{\omega^2+(\omega_{n}\pm i\mu)^2} \, ,
\label{sums}
\end{eqnarray}
and similar expressions for second derivatives.
Here $\omega^2=m_0^2+{\bf p}^2$. Note that a single spatial
derivative would give a zero result, and
that $D_{0}G_{\pm}(0)$ is zero at $\mu=0$. These terms with temporal
covariant derivatives contribute new terms to $P(\mu)$ which are absent
at $\mu=0$. The corresponding temperature dependent finite parts of the
above expressions are
\begin{equation}
\Omega_{\pm}^{(0)}\rightarrow g_{0}(\mu)=-T\int\frac{d^3 p}{(2\pi)^3}
\left[\ln \left( 1-e^{-(\omega+\mu)/T}\right)+\ln \left( 1-
e^{-(\omega-\mu)/T}\right)\right]  \, ,
\label{g0}
\end{equation}
\begin{equation}
G_{\pm}(0)\rightarrow g_{1}(\mu)=
\int\frac{d^3 p}{(2\pi)^3}\frac{1}{2\omega}
\left[\frac{1}{e^{(\omega+\mu)/T}-1}+\frac{1}{e^{(\omega-\mu)/T}-1}\right] \, ,
\label{g1}
\end{equation}
\begin{equation}
D_{0}G_{\pm}(0)\rightarrow
\pm
\frac{1}{2}\frac{\partial g_{0}(\mu)}{\partial\mu}=
\pm\int\frac{d^3 p}{(2\pi)^3}
\left[\frac{1}{e^{(\omega+\mu)/T}-1}-\frac{1}{e^{(\omega-\mu)/T}-1}\right] \, .
\label{g11}
\end{equation}
We also introduce two combinations of $g_{0}(\mu)$ and
$g_{1}(\mu)$,
\begin{eqnarray}
g(\mu)&=&3g_{0}^2(\mu)+3m_{\pi}^2g_{0}(\mu)g_{1}(\mu) \, , \nonumber \\
\bar{g}(\mu)&=&3g_{0}(\mu)g_{0}(0)+\frac{3}{2}m_{\pi}^2
\left(g_{0}(0)g_{1}(\mu)+g_{0}(\mu)g_{1}(0)\right) \, .
\label{g}
\end{eqnarray}
To simplify comparisons with the case $\mu=0$, we follow the notation
of Ref.~\cite{gel} in which the two-loop formula for the pressure
contained functions $g_0$, $g_1$ and their combination
$g=3(g_0^2 +m_{\pi}^{2}g_{0}g_{1})$, so that $g_{0}(0)=g_{0}$,
$g_{1}(0)=g_{1}$ and $g(0)=\bar{g}(0)=g$.
In terms of $g_{0}(\mu)$, $g_{1}(\mu)$, $g(\mu)$ and $\bar{g}(\mu)$
the two-loop pressure at $\mu\neq 0$ takes the following form
\begin{eqnarray}
P(\mu)&=&\frac{1}{2}(g_0 +2g_0 (\mu))
-\frac{m_0^2}{8F^2}
\left(4g_{1}(\mu)g_{1}(0)-g_{1}^2(0)+\frac{2}{m_{\pi}^2}
\left(\frac{\partial g_{0}}{\partial\mu}\right) ^2\right)  \nonumber \\
&+&\frac{m_0^4}{96\pi^2 F^4}(\bar{l}_{1}+2\bar{l}_{2})
\left(3g_{1}^2(0)+4g_{1}(\mu)g_{1}(0)+8g_{1}^2(\mu)\right)  \nonumber \\
&-&\frac{m_0^4}{64\pi^2 F^4}\bar{l}_3
\left(g_{1}^2(0)+2g_{1}^2(\mu)-\frac{1}{2m_{\pi}^2}
\left(\frac{\partial g_{0}}{\partial\mu}\right) ^2\right)  \nonumber \\
&+&\frac{m_0^4}{256\pi^2 F^4}
\left(13g_{1}^{2}(0)+28g_{1}^{2}(\mu)+4g_{1}(\mu)g_{1}(0)-
\frac{5}{m_{\pi}^2}\left(\frac{\partial g_{0}}{\partial\mu}\right)^2
\right)  \nonumber \\
&+&\frac{1}{48\pi^2 F^4}
\left(\bar{l}_{1} \left[g(0)+2g(\mu)\right]+
2\bar{l}_{2}\left[g(0)+3g(\mu)+2\bar{g}(\mu)\right]\right) \nonumber \\
&-&\frac{1}{1152\pi^2 F^4}
(9g(0) +38g(\mu) +40\bar{g}(\mu)) \, .
\label{pmu}
\end{eqnarray}
Here we introduced the renormalized coupling constants \cite{gal}
$\bar{l}_{1},...,\bar{l}_{4}$ through the relation
\begin{equation}
l_{i}=\gamma_{i}\left(\lambda +\frac{1}{32\pi^2}\bar{l}_{i}\right)
\label{ls}
\end{equation}
where $\lambda$ is a logarithmically divergent term and
$\gamma_{1}=1/3$, $\gamma_{2}=2/3$, $\gamma_{3}=-1/2$ and $\gamma_{4}=2$.
We use dimensional regularization since it preserves gauge invariance.
We use the mass renormalization relation \cite{gal}
\begin{equation}
m_{\pi}^2=m_0^2\left( 1-\frac{m_0^2}{32\pi^2 F^2}\bar{l}_{3}\right)
\label{mrenorm}
\end{equation}
so that $g_{0}$, $g_{1}$, $g$ and $\bar{g}$ in Eq.~(\ref{pmu})
are functions of $m_{\pi}^2$.

It is very useful to compare the representation for the pressure $P(\mu)$
obtained with the method of chiral Lagrangians to the result of the virial
expansion in Sect. 2. It is of course evident that the first term in
Eq.~(\ref{pmu}), which is the ideal gas pressure, is exactly the same
as Eq.~(\ref{pideal}).
Using the low temperature expansions of $g_{0}$ and $g_{1}$
\begin{eqnarray}
g_{0}(\mu)&=&2\cosh(\mu\beta)\frac{T^{5/2}m_{\pi}^{3/2}}{(2\pi)^{3/2}}
e^{-m_{\pi}/T}\left( 1+\frac{15}{8}\frac{T}{m_{\pi}}+...\right)
+O(e^{-2m_{\pi}/T})\, , \nonumber \\
g_{1}(\mu)&=&\cosh(\mu\beta)\frac{T^{3/2}m_{\pi}^{1/2}}{(2\pi)^{3/2}}
e^{-m_{\pi}/T}
\left( 1+\frac{3}{8}\frac{T}{m_{\pi}}+...\right)
+O(e^{-2m_{\pi}/T}) \, ,
\label{glowt}
\end{eqnarray}
in Eq.~(\ref{pmu}) the pressure may be written as a series
\begin{equation}
P(\mu)=T(m_{\pi}T/2\pi)^{3/2}\sum_{n=1}^{\infty}B_{n}(\mu)
\exp(-nm_{\pi}/T) \, .
\label{series}
\end{equation}
The $n=1$ term here is the Boltzmann limit of the free gas pressure,
Eq.~(\ref{pideal}),
while the $n=2$ term contains both $O(e^{-m_{\pi/T}})$ corrections
to this limit and the contribution of the interaction,
\begin{equation}
B_{2}(\mu)=B_{2}^{(0)}(\mu)+B_{2}^{int}(\mu, \bar{l}_{i}) \, .
\label{bb}
\end{equation}
The case of a relativistic free gas corresponds to
\begin{equation}
B_{n}(\mu)=B_{n}^{(0)}(\mu)=
(1+2\cosh(\mu\beta))n^{-5/2}
\left(1+\frac{15}{8n}\frac{T}{m_{\pi}}+...\right) \, .
\label{b0}
\end{equation}

On the other hand, one can start from Eq.~(\ref{pint}), take
the standard low momentum ($q<<m_\pi$) representation
for the phase shifts
\begin{equation}
\sin2\delta_{l}^{I}(q)=2q^{2l+1}\,(m_{\pi}^2+q^2)^{-1/2}
\,(a_l^I+q^2 b_l^I+...)
\label{shift}
\end{equation}
and use the
scattering lengths and effective radii calculated in the chiral
perturbation theory \cite{gal} in terms of the couplings $\bar{l}_i$,
\begin{eqnarray}
a_0^0&=&\frac{7m_0^2}{32\pi^2 F^2}
\left(1+\frac{5m_0^2}{84\pi^2 F^2}(\bar{l}_1 +2\bar{l}_2-\frac{9}{10}\bar{l}_3+
\frac{21}{8})+O(m_0^4)\right) \, , \nonumber \\
a_0^2&=&-\frac{m_0^2}{16\pi^2 F^2}
\left(1-\frac{m_0^2}{12\pi^2 F^2}(\bar{l}_1 +2\bar{l}_2+\frac{3}{8})
+O(m_0^4)\right) \, , \nonumber \\
b_0^0&=&\frac{1}{4\pi F^2}
\left( 1+\frac{m_0^2}{12\pi^2 F^2}(2\bar{l}_1 +3\bar{l}_2
-\frac{13}{16})+O(m_0^4)\right) \, , \nonumber \\
b_0^2&=&-\frac{1}{8\pi F^2}
\left( 1-\frac{m_0^2}{12\pi^2 F^2}(\bar{l}_1 +3\bar{l}_2
-\frac{5}{16})+O(m_0^4)\right) \, , \nonumber \\
a_1^1&=&\frac{1}{24\pi F^2}
\left( 1-\frac{m_0^2}{12\pi^2 F^2}(\bar{l}_1 -\bar{l}_2
+\frac{65}{48})+O(m_0^4)\right) \, .
\label{asbs}
\end{eqnarray}
A straightforward check shows that the contributions of interaction
to the pressure agree in the two approaches
and $B_{2}^{int}(\mu, \bar{l}_{i})$ may be written as
\begin{equation}
B_{2}^{int}(\mu)=\left(\frac{2T}{\pi m_{\pi}}\right)^{1/2}
\left(a(\mu)+\frac{3T}{2m_{\pi}}\left[\frac{1}{2} a(\mu)+m_{\pi}^2 b(\mu)
\right]+
O\left(\frac{T^2}{m_{\pi}^2}\right)\right) \, ,
\label{B}
\end{equation}
where
\begin{eqnarray}
a(\mu)&=&2\cosh(2\mu\beta)\,a_0^2 +2\cosh(\mu\beta)\,a_0^2
+a_0^2+a_0^0 \, , \nonumber \\
b(\mu)&=&2\cosh(2\mu\beta)\,b_0^2 +2\cosh(\mu\beta)\,(b_0^2 +3a_1^1)+
b_0^2 +b_0^0+3a_1^1 \, .
\label{amubmu}
\end{eqnarray}
At $\mu =0$ this coincides with the result obtained
in Ref.~\cite{gel}.
It should be noted that the
$\left( \partial g_{0} / \partial\mu \right)^2$
terms in Eq.~(\ref{pmu}) for the pressure
are proportional
to $\sinh^2 (\mu\beta)$. They are absent at $\mu =0$ and
crucial for agreement with the virial result.

The screening mass is finally obtained from Eq.~(\ref{pmu})
using Eq.~(\ref{mel}),
\begin{eqnarray}
m_{el}^2&=&g_{0}^{\prime\prime}(0)-\frac{m_{\pi}^2}{2F^2}
\left(g_{1}^{\prime\prime}(0)g_{1}(0)
+\frac{(g_{0}^{\prime\prime}(0))^2}{m_{\pi}^2}\right) \nonumber \\
&+& \frac{5m_{\pi}^4}{24\pi^2 F^4} \left[
\left(\bar{l}_1 +2\bar{l}_2 -\frac{3}{8}\bar{l}_3 +\frac{9}{8}\right)
g_{1}^{\prime\prime}(0)g_{1}(0) \right. \nonumber \\
& & \left. - \frac{3}{16}\frac{(g_{0}^{\prime\prime}(0))^2}{m_{\pi}^2}
+\left(\bar{l}_1 +4\bar{l}_2 -\frac{29}{24}\right)
\frac{g^{\prime\prime}(0)}{5m_{\pi}^4} \right] \, ,
\label{mscr}
\end{eqnarray}
where $g_{i}^{\prime\prime}(0)\equiv
\partial^2 g_{i}/\partial\mu ^2 (\mu=0)$.  Odd derivatives of $g_i$
with respect to $\mu$ are zero at $\mu = 0$.
Here we eliminated $m_{0}$ in favor of the physical pion mass $m_{\pi}$.
The first term is of course exactly the same as in
Eq.~(\ref{intmel}). All other terms are due to the interaction.
One can check, as in the case of pressure, that they coincide
to relative order $(T/m_{\pi})^{3/2}$ with what
follows from the interaction part of the virial result in
Eq.~(\ref{intmel}) if the approximation of effective radius is used for
the phase shifts.

In numerical calculation we use the central values of the recent
estimates of the couplings $\bar{l}_{i}$ obtained in Ref.~\cite{rigg},
\begin{eqnarray}
\bar{l}_1 &=&-0.62\pm 0.94,~~~\bar{l}_2 =6.28\pm 0.48, \nonumber \\
\bar{l}_3 &=&~~2.9\pm 2.4,~~~\bar{l}_4 =4.3\pm 0.9 ,
\label{expl}
\end{eqnarray}
(the coupling $\bar{l}_4$ relates $F\approx 87$\, MeV to
the physical coupling $F_{\pi}=93$\, MeV\,\cite{gal}).
The results of this calculation are presented in Fig.~3.
The dashed curve represents noninteracting pions.  The chain-dashed
curve includes the contributions of interactions to order $F^{-2}$,
and the solid curve includes also the contributions of order $F^{-4}$.
We have displayed these separately because the chiral perturbation
theory is naturally expressed as an expansion in inverse powers
of $F$.  See Eqs.~(28) and (39).  There is
reasonable agreement with the virial calculation up to temperatures
of around 80 MeV.  Above 100 MeV the chiral perturbation expansion
does not seem convergent.  The order $F^{-4}$ result starts to blow-up.
This can be traced to the basic derivative expansion of the Lagrangian.
For example, the low momentum expansion of the phase shifts in Eq.~(35)
will be inadequate because the average two-pion
collision energy grows with temperature.

It should be noted that in obtaining Eq.~(\ref{pmu}) and Eq.~(\ref{mscr})
we have actually taken into account three--loop diagrams which contain
up to two thermal loops. Due to these thermal loops the final results
are quadratic in the phase space functions $g_0$ and $g_1$.
The third, $T=0$ loop, is responsible for the
renormalisation of the couplings $l_i$. There is, however, one three-loop
diagram which does not factorize into $T=0$ and $T\neq 0$ loops.
This is the eye-type diagram and we did not take it into account.
As was shown in \cite{gel} it gives a contribution to the pressure
which is proportional to $\exp(-m_{\pi}/T)$, but is suppressed by an
additional pre-exponential factor of $T/m_{\pi}$. It contributes
to the $O(T^2/m_{\pi}^2)$ term in Eq.~(\ref{B}) and thus is related to
$q^4$ terms in the expansion Eq.~(\ref{shift}). This contribution
should become important at $T\sim m_{\pi}$ which explains the deviation
from the virial result which uses experimental information on
phase shifts.

\section{Free Bosons on a Lattice}

In principle Monte Carlo simulations of lattice QCD should predict the
properties of strongly interacting matter at all temperatures, including
the low temperature phase of hadrons.  So far little has been
learned about the low temperature phase due to limitations of finite
lattice size and lattice spacing and the difficulty of doing calculations
with light quarks.  Some years ago, singlet and nonsinglet quark number
susceptibilities were computed on a lattice of size $8^3\times4$
\cite{Gott}.  These susceptibilites are linear combinations of the
baryon and electric susceptibilities.  The latter is just the square
of the electric screening mass up to a factor of $e^2$.  Since then
calculations have also been done on a $10^3\times6$ lattice \cite{priv}
and these were used in Ref.~\cite{irv}.  See Fig.~1.  The temperature range
covered by this larger lattice is about 120 to 190 MeV, which is just in the
interesting regime where a crossover from hadron to quark-gluon degrees
of freedom takes place.  When comparing the results of lattice QCD
with continuum field theory calculations it is important to have an
estimate of how important finite lattice size and spacing effects are.
This is true not only of the lattices just mentioned, but also for
lattices to be used in the upcoming Teraflop project; typical lattices
are expected to be $48^3$.  Is this large enough?

To get a handle on this question we consider a system of noninteracting,
massive, charged scalar bosons on a lattice at finite temperature
and chemical potential.  Fermions with a chemical potential have
been studied on the lattice \cite{fermions1,fermions2} but apparently
there are no reports of the boson calculation in the literature.
Since this is a free theory, the partition function can be evaluated
exactly, and from this one can compute the net charge (or number) density
and the electric susceptibility.

We follow here the notation of Creutz \cite{creutz}.  We consider
a Euclidean lattice of size $N_s^3 \times N_t$ with equal lattice spacing
$a$ in the space and time directions.  Roman indices run from 1 to 3,
Greek indices run from 1 to 4, with 4 being the time direction.  A
lattice site is specified by $x_{\nu} = an_{\nu}$.  The integers
$n$ have allowed values
\begin{equation}
-\frac{N_s}{2} < n_i \leq \frac{N_s}{2} \, , \,\,\,\,\,\,\,\,\,
-\frac{N_t}{2} < n_4 \leq \frac{N_t}{2} .
\end{equation}
The temperature is $T = 1/N_ta$ and the physical length of a side
of the 3-dimensional cube is $L = 1/N_sa$.
Letting $\Phi$ denote the complex scalar field, the action is
\begin{equation}
S = a^4 \sum_{(l,n)} \left[ \frac{\Phi^*(l)-\Phi^*(n)}{a}
\right] \left[ \frac{\Phi(l)-\Phi(n)}{a} \right]
+m^2a^4 \sum_n \Phi^*(n)\Phi(n) \, ,
\end{equation}
where the first sum is over all nearest neighbors $(l,n)$.
We shall impose periodic boundary conditions in the spatial directions,
and of course finite temperature requires the fields to be periodic
in the time direction.

We introduce a chemical potential corresponding to the conserved
charge in the same way as one normally introduces an electromagnetic
vector potential \cite{fermions1}.  That is, we make the replacement
$iA_4 \rightarrow iA_4 + \mu$.
This ensures that the chemical potential couples to exactly
the same charge density as the time component of the vector potential.
This maintains gauge invariance.  Thus, in the action, the only term
which changes is the following one.
\begin{equation}
-a^2 \sum_n \left[ \Phi^*({\bf n},n_4+1) e^{a\mu} \Phi({\bf n},n_4)
+ \Phi^*({\bf n},n_4) e^{-a\mu} \Phi({\bf n},n_4+1) \right]
\end{equation}
We can think of this as giving particles (going forward in time) a
chemical potential $\mu$ and antiparticles (going backward in time)
a chemical potential $-\mu$.  This is usually called a link hopping
term.

In order to carry out the functional integration over the fields it
is convenient to express them in terms of their Fourier components.
\begin{equation}
\Phi(n) = \frac{1}{N_s^3 N_t} \sum_k \tilde{\Phi} (k)
\exp{\left[-2\pi i \left( \frac{{\bf k} \cdot {\bf n}}{N_s}
+ \frac{k_4n_4}{N_t} \right) \right]} \, ,
\end{equation}
where the components of the vector $k$ are allowed the same values as
the components of the vector $n$.  Inserting this Fourier decomposition
into the expression for the action, integrating over the lattice sites
with the help of
\begin{equation}
\frac{1}{N_s^3 N_t} \sum_n
\exp{\left[ 2\pi i \left( \frac{{\bf k} \cdot {\bf n}}{N_s}
+ \frac{k_4n_4}{N_t} \right) \right]} = \delta_{k_{\nu},0} \, ,
\end{equation}
we get the action in the form
\begin{equation}
S = \frac{1}{2}\frac{a^4}{N_s^3 N_t} \sum_k D(k) \left[
\tilde{\phi}_1^2(k) + \tilde{\phi}_2^2(k) \right] \, ,
\end{equation}
where $\Phi = (\phi_1 + i\phi_2)/\sqrt{2}$ and the propagator is
\begin{eqnarray}
D(k) & = & m^2 + \frac{2}{a^2}\sum_i \left[ 1 - \cos\left(\frac{2\pi k_i}
{N_s}\right) \right] \nonumber \\
&+& \frac{2}{a^2}\left[ 1 - \frac{1}{2} \left( \exp{ \left\{ a\mu +
\frac{2\pi i k_4}{N_t} \right\} } +
\exp{ \left\{ - a\mu - \frac{2\pi i k_4}{N_t}
\right\} } \right) \right].
\end{eqnarray}
The logarithm of the partition function is, up to an irrelevant
additive constant, given by
\begin{equation}
\ln Z = \ln \int [d\Phi] e^{-S} = -\ln \det \left[ a^2 D(k) \right] \, .
\end{equation}

Let us define a relativistic lattice energy $\epsilon$ by
\begin{equation}
\epsilon^2 = m^2 + \frac{4}{a^2} \sum_i \sin^2\left( \frac{\pi k_i}{N_s}
\right) \, .
\end{equation}
Let us also define the complex variable
\begin{equation}
z = \exp\left\{ \frac{2\pi i k_4}{N_t} \right\} \, .
\end{equation}
Then the partition function can be expressed as
\begin{equation}
\ln Z = - \sum_k \ln \left[ 2 + a^2\epsilon^2 - z e^{a\mu} - z^{-1} e^{-a\mu}
\right] \, .
\end{equation}
It is expedient to differentiate with respect to the chemical potential
before doing the summations.  This gives the net particle number
(or charge) of the system.
\begin{equation}
N = T \frac{\partial \ln Z}{\partial \mu} = aT\sum_k
\frac{z e^{a\mu}-z^{-1}e^{-a\mu}}{2+a^2\epsilon^2-z e^{a\mu}-z^{-1}
e^{-a\mu}} \, .
\end{equation}

We now perform the sum over $k_4$ analytically with the help of the
formula
\begin{equation}
\sum_{k_4} f\left( e^{\frac{2\pi i k_4}{N_t}} \right) =
\frac{N_t}{2\pi i} \int_C \frac{dz}{z} \frac{f(z)}{z^{N_t}-1} \, ,
\end{equation}
where $C$ is any closed contour containing the points satisfying
$z^{N_t} = 1$ and which does not include the origin $z = 0$.
Thus
\begin{equation}
N = \sum_{\bf k} \frac{1}{2\pi i} \int_C \frac{dz}{z}
\frac{f(z)}{z^{N_t}-1}
\end{equation}
where
\begin{equation}
f(z) = \frac{z e^{a\mu}-z^{-1}e^{-a\mu}}{2+a^2\epsilon^2-z e^{a\mu}-z^{-1}
e^{-a\mu}} \, .
\end{equation}
The function $f$ has simple poles at $z = \exp [a(\pm \omega -\mu)]$
where $\omega > 0$ is defined by
\begin{equation}
\sinh\left( \frac{a\omega}{2}\right) = \frac{a\epsilon}{2} \, .
\end{equation}
If one analytically continues from Euclidean space (imaginary time) to
Minkowski space (real time), as is appropriate for obtaining a response
function, then the Matsubara frequency $2\pi k_4 T i \rightarrow p_0$,
where $p_0$ is a real, continuous energy.  The single-particle energies
are determined by the poles of the propagator.  In the limit of a
vanishing chemical potential these poles are located at $p_0 = \pm \omega$
where $\omega$ is as defined in the equation above.
With some rearrangement we can replace the integral over the single
closed contour $C$ with integrals over three disjoint contours
$C_+$, $C_-$ and $C_0$ encircling the two poles of $f(z)$ and the
origin.  The residues are easily evaluated.  Dividing by the volume
we obtain the number density.
\begin{equation}
{\it n} = \frac{1}{(N_s a)^3} \sum_{{\bf k}} \left[
\frac{1}{e^{(\omega - \mu)/T} - 1} -
\frac{1}{e^{(\omega + \mu)/T} - 1} \right]
\end{equation}
This is a familiar form.
It is now easy to integrate {\it n} with respect to $\mu$ to obtain
the partition function.
\begin{equation}
\ln Z = - \sum_{{\bf k}} \left[ \ln\left( 1 - e^{-(\omega - \mu)/T}
\right) + \ln\left( 1 - e^{-(\omega + \mu)/T} \right) \right]
\end{equation}

In the continuum theory Bose-Einstein condensation occurs when the
chemical potential approaches the mass.  On the lattice the condition
is slightly modified.  The number density diverges as $\mu$ approaches
the critical value determined from
\begin{equation}
\sinh\left( \frac{a\mu_{crit}}{2} \right) = \frac{am}{2} \, .
\end{equation}
This allows for any finite number of particles on the lattice even
when the temperature goes to zero.  In that limit, all particles are
concentrated in the zero momentum mode.

Both the partition function and the number density vanish in the
zero temperature limit ($N_t \rightarrow \infty$ at fixed $a$)
so long as $-\mu_{crit} < \mu < \mu_{crit}$.  They both have the
correct limits as the continuum is approached:  $a \rightarrow 0$,
$N_t \rightarrow \infty$, $N_s \rightarrow \infty$ with $N_t \ll N_s$
and $T = 1/N_t a$ fixed.

Now we turn to the electric susceptibility.  Differentiating $n$ with
respect to $\mu$, and setting $\mu = 0$, we get
\begin{equation}
\frac{\partial n}{\partial \mu}({\rm lattice}) =
2\left( \frac{N_t}{N_s} \right)^3 T^2 \sum_{{\bf k}}
\frac{e^{\omega/T}}{\left( e^{\omega/T} - 1\right) ^2} \, .
\end{equation}
This reproduces the correct expression in the continuum limit,
\begin{equation}
\frac{\partial n}{\partial \mu}({\rm continuum}) =
\frac{2}{T} \int \frac{d^3p}{(2\pi)^3}
\frac{e^{\omega/T}}{\left( e^{\omega/T} - 1\right) ^2} \, ,
\end{equation}
where here $\omega = \sqrt{m^2 + {\bf p}^2}$.  The ratio of these
two expressions, lattice/continuum, depends only on the single
dimensionless variable $T/m$ for a given lattice $N_s^3\times N_t$,
since the lattice spacing can be written as $a = 1/N_tT$.

In Fig.~4 we plot lattice/continuum for lattices of size $8^3\times4$
and $10^3\times6$ for which lattice QCD calculations have been done
\cite{Gott,priv}.  (We recall that in the QCD calculations the scale is
set in such a way that the $\rho$ meson mass has its physical value.
The pion is then too heavy, being about 1/2 the $\rho$ mass even for
the larger of the two lattices.)  The susceptibility on the
$10^3\times6$ lattice gets to within 40\% of the continuum value
when $T/m = 0.35$.  It deviates markedly for both lower and higher
temperatures.  As we shall now discuss, the deviation at low temperature
is caused by finite lattice spacing, while the deviation at high
temperature is caused by finite lattice volume.

Apart from the conditions already mentioned which must be satisfied
if the lattice is to approximate the continuum, we have another.
The lattice spacing must be small compared to all physical length
scales.  Thus one must have $a \ll 1/m$.  This is equivalent to
the condition $1/N_t \ll T/m$.  Therefore the departure of the
lattice susceptibility from the continuum limit will be greater
and greater as the temperature gets smaller.  This is seen in the
figure.  From the expressions given above we can readily evaluate
the susceptibilities in the low temperature limit.
\begin{displaymath}
\frac{\partial n}{\partial \mu}({\rm lattice})
\rightarrow 2N_t^5 \frac{T^4}{m^2} \, ,
\end{displaymath}
\begin{displaymath}
\frac{\partial n}{\partial \mu}({\rm continuum})
\rightarrow \frac{1}{4T} \left(\frac{2mT}{\pi}\right)^{3/2} e^{-m/T} \, ,
\end{displaymath}
\begin{equation}
{\rm lattice/continuum} = (2\pi)^{3/2} N_t^5 \left( \frac{T}{m} \right)^{7/2}
e^{m/T} \, .
\end{equation}
The ratio diverges exponentially as $T/m \rightarrow 0$.  To get
accurate results we obviously cannot go too low in $T/m$ for a fixed
value of $N_t$.

The high temperature limit is equivalent to letting the mass go to
zero.  The susceptibility of the lattice diverges as the mass
vanishes because of the zero momentum mode.  This is not true of
the susceptibility in the continuum; in the continuum the integral
is convergent in the infrared because of the factor $p^2dp$.
Therefore the lattice/continuum ratio also increases at large
values of $T/m$.  However, this is a finite lattice volume effect,
not finite lattice spacing effect.
We can see it in the following
ways.  If we consider a box of volume $L^3$ instead of the continuum
limit then we would make the replacement
\begin{equation}
\int \frac{d^3p}{(2\pi)^3} \rightarrow \frac{1}{L^3} \sum_{{\bf p}} \, .
\end{equation}
Then the susceptibility diverges in the zero mass limit because
of the ${\bf p} = {\bf 0}$ mode, which was suppressed in the integral.
Numerically we can see this if we increase the ratio $N_t/N_s$.
To approach the zero lattice spacing, infinite volume, limit we
require (among other conditions) that this ratio be small in
order that many thermal wavelengths fit within the box.  That is,
$1/T \ll L$.  In this sense the lattice $8^3\times4$ is `bigger'
than the lattice $10^3\times6$, as is apparent in Fig.~4.

In Fig.~5 we plot the lattice/continuum ratio for lattices of size
$48^3$ by 12, 16 and 24.  These may be typical for the upcoming
Teraflop project.  On the whole the susceptibilities are much
closer to the continuum values than was true for the smaller
lattices.  However, the small and large $T/m$ behavior is
still apparent, as discussed above.  The condition
$1 \ll N_t \ll N_s$ is also apparent in the figure.

Since hadrons with masses in the range of 140 to 1000 MeV and beyond
are important for the electric screening mass in the temperature
range of 50 to 170 MeV or so, it is clear that very large lattices
are required to at least reproduce the noninteracting gas results.
Whether interaction effects, and the composite nature of the hadrons,
are more or less sensitive to finite lattice spacing and volume
is not known.  It is known, and obvious, that any interaction effects
that get big contributions from long wavelength modes are even
more sensitive to finite lattice volume \cite{elze}.

\section{Conclusion}

To get insight into the nature of the expected phase transition
(or rapid crossover) of hot
hadronic matter to quark-gluon plasma, it is instructive to
investigate the temperature dependence of different quantities
characterizing the system on both sides of the transition point (or
crossover region).
We have considered here one such quantity, the electric screening mass,
in the case of hadronic gas. We used two different methods:
relativistic virial expansion and chiral perturbation theory. They
give very similar results up to $T \sim 80$ MeV.  At high temperature
the applicability of both approaches become doubtful, for somewhat
different reasons.
In the virial expansion, multi-pion interactions would have to be taken
into account.  This is difficult to do theoretically, apart from
the fact that multi-pion scattering amplitudes are practically
impossible to obtain experimentally.
The advantage of the chiral perturbation expansion is that it very
generally takes into account the basic symmetries of QCD in an
expansion in powers of the pion mass and space-time derivatives.
This is also its limitation.  The $\rho$ meson cannot be generated in
any finite order; it can arise only from an infinite series of
derivative terms in the effective Lagrangian.  As the temperature
rises, the average relative energy in hadron collisions also rises.
Therefore, more and more derivative terms in the expansion must
be kept.  Not only is this difficult to do theoretically, but
the coefficients of the higher order terms are not known yet from
phenomenology.  The application of chiral perturbation theory, at
least in its present form, is also limited to temperatures below
the pion mass.

Numerical simulations on the lattice provide a straightforward
possibility to go all the way from low to high temperatures.
However, this approach has intrinsic problems due to finite
lattice spacing and lattice volume effects.
We have shown here that to reproduce the free gas results for the
screening mass would require very large lattices. Analytic methods,
such as the chiral perturbation theory and the virial expansion, should
be useful to obtain estimates on the magnitude of the final volume
effects for the interacting hadron gas \cite{ep}.

\vspace{.2in}

\noindent $^{\dagger}$ On leave from Institute of Theoretical and
Experimental Physics, Moscow 117259, Russia.

\section*{Acknowledgements}

We would like to thank H. Leutwyler, L. McLerran and M. Stone
for helpful conversations.
This work was supported by the Schweizerischer Nationalfonds,
by the Minnesota Supercomputer Institute, and
by the U.S. Department of Energy under grant DOE/DE-FG02-87ER40328.

\newpage
\section*{Figure Captions}

Figure 1: The square of the electic mass in units of $e^2T^2$ vs.
temperature.  At low $T$ the two lines represent the contributions
from pions, and pions plus $\rho$ mesons.  At high $T$ the three lines
represent the contributions from up and down quarks computed to
the indicated order in the QCD coupling.  The data points are from
lattice QCD calculations on a $10^3\times6$ lattice.  For free massless
up and down quarks on a lattice of this size $m_{el}^2/e^2T^2 = 35/36$
as indicated by the dashed line in the upper right-hand corner.
In this figure the pion mass was taken to be one-half the $\rho$ mass
in order to facilitate comparison with the lattice results.  Taken
from Ref.~\cite{irv}.

\vspace{.25in}

\noindent Figure 2:  Square of electric screening mass in units of
$e^2T^2$ vs. temperature. The dashed line is the contribution from a
free gas of $\pi$ mesons; the chain-dashed line includes free $\rho$
mesons as well. The solid line is the square of the screening mass
for an interacting pion gas computed using empirical $\pi\pi$ phase
shifts in the relativistic virial expansion.

\vspace{.25in}

\noindent Figure 3:  Same as Fig. 2 but from chiral perturbation theory.
The dashed line represents free pions.  The chain-dashed line includes
the effects of interactions to order $F^{-2}$.  The solid line includes
the effects of interactions to order $F^{-4}$.

\vspace{.25in}

\noindent Figure 4:  Ratio of $\partial n/\partial \mu$ on the lattice
of specified size to the continuum as a function of temperature over
mass.

\vspace{.25in}

\noindent Figure 5:  Ratio of $\partial n/\partial \mu$ on the lattice
of specified size to the continuum as a function of temperature over
mass.

\end{document}